\documentclass[a4paper,12pt]{article}
\usepackage{amsmath}
\usepackage{amssymb}
\usepackage{amsfonts}
\usepackage{graphics}
\usepackage{epsfig}

\begin{document}
\renewcommand{\theequation}{\thesection.\arabic{equation}}
\thispagestyle{empty}
\vspace*{-1.5cm}

\begin{center}
{\Large\bf  LOCAL INTERACTIONS OF HIGHER-SPIN POTENTIALS THAT ARE GAUGE
INVARIANT IN LINEAR APPROXIMATION} \\
\vspace{4cm}
{\large Werner R\"uhl}\\
Department of Physics, Technical University of Kaiserslautern\\P.O.Box 3049,
67653 Kaiserslautern, Germany \\
\vspace{5cm}
\begin{abstract}
We study connected Wightman functions of $N$ conserved currents, each of which is formed from a scalar field and has even spin $l_{i}$. The UV divergence of this vertex function is regularized by the analytic continuation in the space dimension $D\longrightarrow  D-\epsilon$. We evaluate the residue of $\epsilon ^{-1} $ only, which is a local interaction Lagrangian density and gauge invariant in linear approximation.
\end{abstract}
\vspace{5cm}
{\it November 2008}
\end{center}
\newpage

\section{Introduction}
Local gauge invariant interactions of massless higher-spin fields $h^{(l)}(z)$ 
that are symmetric tensors of rank (or "spin") $l$ are the starting point for any analysis of the quantum field structure or possible physical applications of this theory. It is quite remarkable that knowledge of explicit interaction Lagrangians is quite limited. The argument $z$ of these fields range over 
Minkowski or $\mathbf{AdS}$ space and by contraction with $l$ tangent vectors $a$ of such space at point $z$ we can write these tensor fields as $h^{(l)}(z;a)$. 

Let us first consider a field theory with one such field of fixed spin $l$.
Then its free field Lagrangian $\mathcal{L}_{0} [h^{(l)}]$ satisfies
\begin{equation}
\frac{\delta\mathcal{L}_{0}}{\delta  h^{(l)}(z;a)} = \mathcal{F} h^{(l)}(z;a)
\end{equation}
where $\mathcal{F}$ is the Fronsdal operator \cite{Frons}. It is gauge invariant 
under 
\begin{equation}
\delta_{0}h^{(l)}(z;a) = (a\nabla) \varepsilon^{(l-1)}(z;a) 
\end{equation}
with a gauge function $\varepsilon$ that is a symmetric tensor of rank $l-1$.
The trace constraints
\begin{equation}
\Box^{2}_{a} h^{(l)}(z;a) = \Box_{a} \varepsilon^{(l-1)}(z;a) = 0
\end{equation}
are always (on and off shell) assumed to hold. 

We want to construct an interacting spin $l$ field theory which contains only one (universal) coupling constant $\kappa$ which is connected with gauge invariance. An ansatz for the complete Lagrangian density of such theory could be 
\begin{equation}
L_{c} = L_{0} +\kappa L_{1}+ \kappa^{2} L_{2} +...+\kappa^{N} L_{N}, 
\mathcal{L} = \int L_{c}
\end{equation}
Let the number of derivatives in $L_{n}$ be $k_{n}$ and the polynomial order in the fields $h^{(l)}$ be $n+2$. Then we can $\kappa^{n}$ let compensate the dimension of the fields. This forces us to introduce another length scale $R$
to compensate for the derivatives. In the case of the $\mathbf{AdS}$ field theory this could be the $\mathbf{AdS}$ radius $L$. 

We must now complete the gauge transformation in the same fashion 
\begin{equation}
\delta h^{(l)} = \delta_{0} h^{(l)} + \kappa \delta_{1} h^{(l)} + \kappa^{2}
\delta_{2} h^{(l)} + ...
\end{equation}
expanding as far as is necessary. Each $\delta_{n} h^{(l)}$ is assumed to be linear in $\varepsilon^{(l-1)}$ and of polynomial order $n$ in $h^{(l)}$. The 
constraint that $\mathcal{L}$ is invariant under this gauge transformation 
(which is some kind of current conservation on shell) can be solved recursively in powers of $\kappa$. The solution depends on some free parameters $\{c_{i}\}$. Applying a field redefinition 
\begin{equation}
h^{(l)}_{new} = h^{(l)} + \sum_{n\geq 1} \kappa^{n} \chi_{n}
\end{equation}
where $\chi_{n}$ are local polynomials of order $n+1$ of $h^{(l)}$ and its derivatives, allows to eliminate some of them. In addition we may postulate that the commutators of two (infinitesimal) gauge transformations satisfy the Jacobi identity. In this case the remaining $\{c_{i}\}$ take definite values and the solution of the interaction Lagrangian densities becomes unique. 
At the same time the gauge transformations, eventually taking the shape of
formal power series, become unique, too.

Following \cite{Ber} who study the cases $l=2$
and $l=3$ explicitly and formulate the general algorithm in some detail,
we propose that the following assertions are true in general:
\begin{enumerate}
\item For even $l$ a single field $h^{(l)}$ possesses a unique Lagrangian density and gauge group for any given $L_1$;
\item For odd $l$ a multiplet of $N\geq2$ fields $h^{(l)}_{i}, i\in 
\{1,2,3.. N\}$ possesses a unique Lagrangian density and gauge group for any given $L_1$ with a 
totally antisymmetric numerical factor $f_{ijk}$ multiplying $h^{(l)}_{i},h^{(l)}_{j}, h^{(l)}_{k}$ in $L_{1}$, and the gauge group involves a Lie group factor;
\item The minimal number of derivatives in $L_{1}$ is $l$;
\item For this "minimal" model the number of derivatives in $\delta_{1}h^{(l)}$
is $l-2$ and these apply to both $ \varepsilon^{(l-1)}$ and $h^{(l)}$.
\end{enumerate}

It is obvious and trivial that some $L_{1}$ can be constructed from three curvatures of $h^{(l)}$ \cite{MR2}
and that this form of $L_{1}$ contains $3l$ derivatives. Whether it is possible to construct $L_{1}$ from two connections and one curvature is not obvious.
This would amount to the occurrence of $3l-2$ derivatives. This is exactly the number of derivatives we will find (for $D=4$) in the construction of the main part of this work.

\setcounter{equation}{0}
\section{The three-point vertex function}
In \cite{MR} the scalar one-loop correction to the propagator of the higher-spin field $h^{(l)}$ in $\mathbf{AdS}_{D}$ space was analyzed with the aim to isolate the anomaly that it produces. We started from regularizing the UV divergence of this loop function by an analytic continuation $D \longrightarrow D-\epsilon$
and extracting the singular term proportional to $\epsilon^{-1}$. Its residue is proportional to a local linearly gauge invariant Green function. Here we want to generalize this method starting from an arbitrary connected $N$-point Green function for even $l_{i}$ only
\begin{equation}
\Gamma^{(N)} = < \prod_{i=1}^{N}J^{(l_{i})}(z_{i}; a_{i})>_{conn}
\end{equation}
In \cite{MR} the conserved and traceless currents were constructed beginning with a germ current $j^{(l)}(x;a)$ on flat space (Minkowski space) in which trace terms proportional to a nonzero power of $a^{2}$ are neglected. The full $\mathbf
{AdS}_{D}$ current $J^{(l)}$ can be recovered from the germ $j^{(l)}$ by a simple algorithm displayed in \cite{MR}, eqs. (1)-(8). Thus in this work we
will calculate the Green function, say
\begin{equation}
\Gamma^{(3)} = <J^{(l_1)}(x_1;a) J^{(l_2)}(x_2;b) J^{(l_3)}(x_3;c)>_{conn}
\end{equation}
neglecting trace terms containing $a^{2}, b^{2}, c^{2}$ (but keeping $(ab), (bc), (ca)$) and the noncommutativity of gradients that would result in terms proportional to negative powers of the space curvature $L$. Using the techniques developed in \cite{MR} and in other papers of these authors it is possible to derive complete expressions.

With the scalar field $\sigma(z)$ on AdS$_{D}$ space we construct the current
\begin{equation}
\frac{1}{2} \sum_{p=0}^{l} A_{p }(a\nabla)^{l-p}\sigma(z)(a\nabla)^{p}\sigma(z)
+ \textnormal{trace and space curvature terms}
\end{equation}
with
\begin{equation}
A_{p} = (-1)^{p}\frac{{l\choose p}{ l+D/2-2 \choose p}}{{p+D/2-2 \choose p}}
\end{equation}
for which (for even $l$) 
\begin{equation}
A_{p} = A_{l-p}
\end{equation}
and for $D=4$ 
\begin{equation}
A_{p} = (-1)^{p}{l\choose p}^{2}
\end{equation}

In order to evaluate the Green function (2.2) for free sigma fields we apply 
Wick's theorem with the AdS propagators 
\begin{equation}
F(\zeta_{12}) = <\sigma(z_1)\sigma(z_2)> = \frac{1}{8\pi^{2}} 
(\frac {1}{\zeta_{12} -1} +\frac{1}{\zeta_{12} +1})
\end{equation}
and $\zeta_{12}$ is $cosh \eta_{12}$, $\eta_{12}$ the geodesic angle between the two points $z_1$ and $z_2$. We obtain eight contractions that are all equal
and together we get
\begin{eqnarray}
\Gamma^{(3)} = \sum_{p_1,p_2,p_3}A_{p_1}^{(l_1)}A_{p_2}^{(l_2)}A_{p_3}^{(l_3)}
(a\nabla_1)^{p_1}(b\nabla_2)^{l_2-p_2}F(\zeta_{12})\qquad\qquad\nonumber\\(a\nabla_1)^{l_1-p_1}
(c\nabla_3)^{p_3}F(\zeta_{31})(b\nabla_2)^{p_2}(c\nabla_3)^{l_3-p_3}F(\zeta_{23}) \qquad\qquad \nonumber\\ + \textnormal{trace and space curvature terms}\qquad\qquad\qquad
\end{eqnarray}

In order to perform the differentiations we introduce the shorthands
\begin{eqnarray}
I(1_{a},2_{b}) = (a\nabla_1)(b\nabla_2)\zeta_{12}\\
I(1_{a}, 2) = (a\nabla_1)\zeta_{12}\\
I(1, 2_{b}) = (b\nabla_2)\zeta_{12}
\end{eqnarray}
etc. and use formulas of \cite{MR}, Appendix A
\begin{eqnarray}
(a\nabla_1)I(1_a,2) = a^{2} \zeta_{12}\\
(a\nabla_1)I(1_a,2_b) = a^{2} I(1,2_b)
\end{eqnarray}
Denoting
\begin{equation}
F^{(n)} = (\frac{d}{d\zeta})^n F(\zeta)
\end{equation}
we obtain for the differentiations 
\begin{eqnarray}
(a\nabla_1)^{p_1}(b\nabla_2)^{l_2-p_2} F(\zeta_{12}) = \qquad\qquad\qquad\qquad\qquad \nonumber\\
\sum_{n_1=0}^{l_2-p_2} C_{p_1p_2n_1}^{(l_2)} I(1_a,2)^{p_1+p_2-l_2+n_1}
I(1,2_b)^{n_1}I(1_a,2_b)^{l_2-p_2-n_1} F^{(p_1+n_1)}(\zeta_{12}) +
\nonumber \\
+ \textnormal{trace and space curvature terms}\qquad\qquad\qquad
\end{eqnarray}
where 
\begin{equation}
C_{p_1p_2n_1}^{(l_2)} = \frac{(l_2-p_2)!}{n_1^!} {p_1 \choose p_1+p_2+n_1-l_2} 
\end{equation}
and the labels run over the intervals
\begin{eqnarray}
0\leq p_{1,2,3}\leq l_{1,2,3} \qquad\qquad \\
max\{0, l_2-p_1-p_2\}\leq n_1 \leq l_2-p_2
\end{eqnarray}
etc. Inserting three factors of type (2.15) into (2.8), we obtain a special 
tritensor expressed by an algebraic tensor basis of nine elements $I$ as defined in (2.9)-(2.11) but depending on three pairs of coordinates $\zeta_{12}, \zeta_{23}, \zeta_{31}$. For the construction presented in this work only the first term (2.7) of the scalar field propagator is used which is singular at the two points coinciding.

\setcounter{equation}{0}
\section{Extracting the leading singularity}

Restricting the analysis to the terms that survive in the flat space limit 
(which is the same as neglecting the space curvature terms) we can replace
the singular function
\begin{equation} 
S(z_1,z_2,z_3) = (\zeta_{12}-1)^{-\lambda}(\zeta_{23}-1)^{-\mu}(\zeta_{31}-1)
^{-\nu}
\end{equation}
by the analogous expression on flat space $\mathcal{R}_{D}$
\begin{equation}
\mathcal{F}(x_1,x_2,x_3) = ((x_1-x_2)^{2})^{-\lambda}((x_2-x_3)^{2})^{-\mu}
((x_3-x_1)^{2})^{-\nu}
\end{equation}
whose singular properties are revealed by a Fourier transformation
\begin{eqnarray}
\mathcal{G}(p_1,p_2,p_3) = \int dx_1dx_2dx_3 \mathcal{F}(x_1,x_2,x_3) \exp i\sum_{i=1}^{3}
x_{i}p_{i} \nonumber\\= \delta(p_1+p_2+p_3)\Phi(p_1,p_2)\qquad\qquad\qquad
\end{eqnarray}
Using that 
\begin{eqnarray}
(x^2)^{-\lambda} = C_{\lambda} \int dq e^{-ixq}(q^2)^{\lambda-D/2}\\
C_{\lambda}^{-1} = 2^{2\lambda} \pi^{D/2} \frac{\Gamma(\lambda)}{\Gamma(-\lambda +D/2)}
\end{eqnarray}
we obtain 
\begin{eqnarray}
\mathcal{G}(p_1,p_2,p_3) = C_{\lambda}C_{\mu}C_{\nu}(2\pi)^{3D} \int dq_1dq_2dq_3
(q_{1}^{2})^{\lambda -D/2}(q_{2}^{2})^{\mu-D/2}(q_{3}^{2})^{\nu-D/2}\nonumber\\
\delta(p_1-q_1+q_3)\delta(p_2-q_2+q_1) \delta(p_3-q_3+q_2)\qquad\qquad\qquad
\end{eqnarray}
Applying standard methods for evaluation of Feynman integrals, in the present case for conformal field theories, we derive
\begin{eqnarray}
&&\Phi(p_1,p_2) = (2\pi^{2})^{D}2^{2(D-\lambda-\mu-\nu)}\frac{\Gamma(D-\lambda-\mu-\nu)}{\Gamma(\lambda)\Gamma(\mu)\Gamma(\nu)}\
\int_{s_{i} \geq 0}ds_1ds_2ds_3\delta(1-s_1-s_2-s_3)\nonumber\\ && \times s_1^{-\lambda+D/2-1}s_2^{-\mu +D/2-1}s_3^{-\nu+D/2-1}  \{s_1s_2p_2^{2}+s_1s_3p_1^{2}
+s_2s_3p_3^{2}\}^{\lambda+\mu+\nu-D} 
\end{eqnarray}

Now we make use of the fact that $\lambda, \mu, \nu$ are natural numbers
and apply the conventional method of dimensional regularization replacing
\begin{equation}
D \longrightarrow D - \epsilon
\end{equation}
maintaining an integer value for $D$. We recognize that there exists only one series of first order poles whenever
\begin{equation}
m= \lambda +\mu +\nu -D \in \mathbf{N}_0
\end{equation}
Then for $\epsilon$ tending to zero and using
\begin{equation}
\Gamma(-m-\epsilon) = \Gamma(m+1+\epsilon)^{-1}\frac{\pi}{\sin\pi(m+1+\epsilon)}  
= \frac{(-1)^{m+1}}{m!} \epsilon^{-1} +O(1)
\end{equation}
we obtain as residue of $\Phi(p_1,p_2)$ in $\epsilon$
\begin{eqnarray}
\Omega_{m}\int ds_1ds_2ds_3 \delta(s_1+s_2+s_3-1) s_1^{-\lambda+D/2-1} s_2^{-\mu +D/2-1} s_3^{-\nu+D/2-1} \nonumber\\ \times \{s_1s_3p_1^{2} + s_1s_2p_2^{2}+s_2s_3p_3^{2}\}^{m} \mid _{p_3=-p_1-p_2}
\end{eqnarray}
where
\begin{equation}
\Omega_{m} =\frac{(-1)^{m+1}}{m!} 2^{-2m}(2\pi^2)^{D}\{\Gamma(\lambda)\Gamma(\mu)
\Gamma(\nu)\}^{-1}
\end{equation}

From (3.11) we inspect that the residue is a local distribution (understood in coordinate space as usual in QFT). If we denote the integral by $P_{m}(p_1^{2},p_2^{2},p_3^{2})$, understanding that it is a homogeneous  
polynomial in the three momentum squares of degree m, the inverse Fourier transform is indeed
\begin{eqnarray}
(2\pi)^{-3D} \Omega_{m} \int dp_1dp_2dp_3 \delta(p_1+p_2+p_3) P_{m}(p_1^2,p_2^2,
p_3^2)\exp(-i\sum x_{i}p_{i})\nonumber\\
= (2\pi)^{-D}\Omega_{m}P_{m}(-\Box_1,-\Box_2,-\Box_3) \delta(x_1-x_3)\delta(x_2-x_3)
\end{eqnarray}
The integral in (3.11) is easily evaluated and gives 
\begin{equation}
P_{m}(p_1^2,p_2^2,p_3^2) = \sum_{r_1r_2r_3} R_{r_1r_2r_3}^{(m)}
(p_1^2)^{r_1}(p_2^2)^{r_2}(p_3^2)^{r_3}
\end{equation}
and the rational numbers $R$ can be expressed by
\begin{eqnarray}
&&R^{(m)}_{r_1r_2r_3} = \delta_{m,r_1+r_2+r_3}{ m \choose r_1,r_2,r_3 } \nonumber\\ &&\frac{\Gamma(r_1+r_2-\lambda+D/2)\Gamma(r_2+r_3-\mu +D/2)\Gamma(r_3+r_1-\nu+D/2)}{\Gamma(2m-\lambda-\mu-\nu+3D/2)}
\end{eqnarray}

\setcounter{equation}{0}
\section{The N-vertex function}

An analogous approach can be formulated for an $N$-vertex function $\Gamma^{(N)}$ which couüples $N$ higher spin fields to a loop of the $\sigma$-field 
\begin{equation}
\Gamma^{(N)} = <\prod_{i=1}^{N} J^{(l_{i})}(z_{i}; a_{i})>_{conn}
\end{equation}
Applying Wick's theorem we obtain contributions from $\frac{1}{2}(N-1)!$
graphs. One of them contains the maximally singular term
\begin{equation}
S^{(N)} = \prod_{i=1}^{N}(\zeta_{i,i+1}-1)^{-\mu_{i}} 
\end{equation}
Other graphs contain different cycles of the variables $\{\zeta_{ij}\}$
and can be treated the same way.

By the same arguments as for the case $N=3$ we consider the planar analogue
(see (3.2)-(3.6)) which yields the Fourier transform
\begin{eqnarray}
\mathcal{G}^{(N)}(p_1,p_2,p_3,\dots p_{N}) = (2\pi)^{ND}\prod_{i=1}^{N} C_{\mu_{i}}
\int dq_1dq_2dq_3...dq_{N}\prod_{i=1}^{N}(q_{i}^{2})^{\mu_{i}-D/2}\nonumber\\ 
\delta(p_1-q_1+q_{N})\delta(p_2-q_2+q_1)\dots \delta(p_{N}-q_{N}+q_{N-1})\qquad\qquad\\
=\delta(\sum_{i=1}^{N}p_{i})\Phi^{(N)}(p_1,p_2,\dots p_{N-1})\qquad\qquad\qquad\qquad\\
\Phi^{(N)}(p_1,p_2,\dots p_{N-1}) = 2^{ND-2\sum\mu_{i}}\pi^{(N+1)D/2}\frac{\Gamma((N-1)D/2 -\sum \mu_{i})}{\prod_{i}\Gamma(\mu_{i})} \nonumber\\
\int ds_1ds_2\dots ds_{N} \delta(1-\sum _{i}s_{i})\{\sum_{ij}A_{ij}(s) (w_{i}w_{j})\}^{\sum\mu_{i} -(N-1)D/2}\qquad
\end{eqnarray}
where 
\begin{eqnarray}
A_{ij}(s) =s_{i}\delta_{ij} -s_{i}s_{j} \\
w_{i} =\sum_{j=1}^{i} p_{j} \qquad (w_{N} = 0)
\end{eqnarray}
Again we obtain a unique series of first order poles at
\begin{equation}
m = \sum_{i} \mu_{i} -(N-1)D/2 \in \bf{N}_0
\end{equation} 
and the residues are proportional to
\begin{equation}
\int_{\mathcal D} ds_1ds_2\dots ds_{N-1}\delta(1-\sum s_{i})\{\sum_{ij}
A_{ij}(s)(w_{i}w_{j})\}^{m}
\end{equation}
By a change of integration variables we can turn the integral (4.9) into
\begin{equation}
\int_{\mathcal D}ds_1 ds_2\dots ds_{N-1} \{\sum_{i\leq j}B_{ij}(s)(p_{i}p_{j})\}^{m}
\end{equation}
where
\begin{eqnarray}
B_{ij}(s) = s_{j}(1-s_{i})(2-\delta_{ij})\qquad\\
\mathcal D =\{1\geq s_1\geq s_2\dots \geq s_{N-1} \geq 0\}
\end{eqnarray}
From (4.8) we inspect that for odd $D$ we must have odd $N$ to obtain a UV singularity.

After Fourier transformation we obtain an invariant differential operator of the order $2m$ with rational coefficients which is applied to the local distribution
\begin{equation}
\prod_{1 \leq i \leq N-1} \delta(x_{i} - x_{N})
\end{equation}

\setcounter{equation}{0}
\section{The residue of $\Gamma^{(3)}$}

The final aim is clear: The singular functions studied in the preceding two sections must be inserted into the $\epsilon$-residue of $\Gamma^{(N)}$ taking
into account that the tensorial structure of the Green functions contributes polynomials in the coordinates that reduce the singularities. Using coordinates from the flat tangential space at, say, $x_{N}$ in the $\mathbf{AdS}$ space (respectively the flat space coordinates themselves for $\mathbf{R}_{D}$) we expect an expansion of the form
\begin{equation}
Res_{\epsilon} \Gamma^{(N)} = \{\sum_{k=0}^{k_{max}} \mathcal W^{(N)}_{k} (x;a) 
   L^{-2k}\} \prod_{i=1}^{N-1} \delta(x_{i} -x_{N})
\end{equation}
Each $\mathcal W^{(N)}_{k}$ is a differential operator defined at $L\rightarrow \infty$ and, if the currents are tracefree, their parts polynomially dependent on $a_1^2,a_2^2,\dots a_{N}^2$ can be neglercted first because there is an elementary algorithm to reconstruct them.
On the other hand we know from the introduction that for $N=3$ at least one 
gauge invariant Lagrangian density of fields $h^{(l_1)}, h^{(l_2)}, h^{(l_3)}$
exists with a minimal number of derivatives $\Delta_{min}$. From the 
above expansion (say for $N=3$) we speculate that a whole basis $V^{(N)}_{\Delta}$ of such forms exists with degree $\Delta$ of derivatives which range over $\Delta_{max} \geq \Delta \geq \Delta_{min}$. Now by dimensional analysis we conclude that this degree ("$deg$") satisfies
\begin{equation}
deg\{W^{(N)}_{k}\} = \Delta_{max} -2k
\end{equation}
and moreover that each $W^{(N)}_{k}$ can be expanded in the basis as
\begin{equation}
W^{(N)}_{k} = \sum_{n\geq 0} \alpha_{k,n} V^{(N)}_{k+n} L^{-2n}
\end{equation}
Calculation of the residue of $\Gamma^{(N)}$ would thus enable us to derive the basis $\{V^{(N)}_{\Delta}\}$ completely by linear algebra. 

We want to be more explicit and study $\Gamma^{(3)}$ in some detail (in $\mathbf{AdS}$ notation).
From (2.15) we obtain
\begin{eqnarray}
\Gamma^{(3)} = \sum_{p_1p_2p_3} A^{(l_1)}_{p_1}A^{(l_2)}_{p_2}A^{(l_3)}_{p_3}
\sum_{n_1n_2n_3} C_{p_1p_2n_1}^{(l_2)}C_{p_2p_3n_2}^{(l_3)} C_{p_3p_1n_3}^{(l_1)}\qquad\qquad \qquad \qquad\nonumber\\
\times I(1_{a},2)^{p_1+p_2-l_2+n_1}I(1,2_{b})^{n_1}I(1_{a},2_{b})^
{l_2-p_2-n_1}I(2_{b},3)^{p_2+p_3-l_3+n_2}\nonumber\\ \times I(2,3_{c})^{n_2}I(2_{b},3_{c})^{l_3-p_3-n_2} 
 I(3_{c},1)^{p_3+p_1-l_1+n_3}I(3,1_{a})^{n_3}\qquad\qquad\nonumber\\ \times I(3_{c},1_{a})^{l_1-p_1-n_3}
F^{(p_1+n_1)}(\zeta_{12})F^{(p_2+n_2)}(\zeta_{23})F^{(p_3+n_3)}(\zeta_{31})\qquad
\end{eqnarray}
The last factor is in explicit form 
\begin{eqnarray}
F^{(p_1+n_1)}(\zeta_{12})F^{(p_2+n_2)}(\zeta_{23})F^{(p_3+n_3)}(\zeta_{31}) =
 (2\pi)^{-6} (-1)^{\sum (p_{i}+n_{i})}\prod_{i}(p_{i}+n_{i})!\nonumber\\ \times
(\zeta_{12}-1)^{-p_1-n_1-1}(\zeta_{23}-1)^{-p_2-n_2-1}(\zeta_{31}-1)^{-p_3-n_3-1}\qquad\qquad
\end{eqnarray}
So we get from (3.9) 
\begin{equation}
m= \sum_{i}(p_{i}+n_{i}) +3-D
\end{equation}
From the coefficients $C_{p_1p_2n_1}^{(l_2)}$ etc. we get then the bound
\begin{equation}
m\leq l_1+l_2+l_3 +3-D
\end{equation}

Now we limit our discussion to the flat case, namely the term $k=0$ in (5.1). 
Though constant factors are not very relevant, let us be precise for the moment. We obtain
\begin{equation}
\zeta_{12} -1 = u_{12} \longrightarrow \frac{1}{2}(x_1-x_2)^2 = \frac{1}{2}(x_{12})^2
\end{equation}
which reflects the fact that $u$ is half the square of the chordal distance.
In the same way we get
\begin{eqnarray}
I(1_{a},2_{b})\longrightarrow -a\cdot b\\
I(1_{a},2) \longrightarrow a\cdot x_{12} \\
I(1,2_{b}) \longrightarrow -b \cdot x_{12}
\end{eqnarray}
The $I$-factors in (5.4) give
\begin{eqnarray}
(a\cdot x_{12})^{u_1}(-b\cdot x_{12})^{n_1}(-a \cdot b)^{w_1}(b \cdot x_{23})^{u_2}(-c \cdot x_{23})^{n_2}\nonumber\\(-b \cdot c)^{w_2} (c \cdot x_{31})^{u_3}
(-a \cdot x_{31})^{n_3} (-c \cdot a)^{w_3}\qquad
\end{eqnarray}
where we chose a cyclic notation
\begin{eqnarray}
 u_1=p_1+p_2-l_2+n_1,\quad w_1= l_2-p_2-n_1 \\
 u_2=p_2+p_3-l_3+n_2,\quad w_2=l_3-p_3-n_2\\
 u_3=p_3+p_1-l_1+n_3,\quad w_3=l_1-p_1-n_3
 \end{eqnarray}
The distribution part takes at the end the form as the $r_1,r_2,r_3$ sum over
\begin{eqnarray}
\int dx_1dx_2dx_3 \delta(x_{12})\delta(x_{13})\Box_1^{r_1}\Box_2^{r_2}\Box_3^{r_3} (a\cdot x_{12})^{u_1} \dots
(-c \cdot a)^{w_3} \nonumber\\ \ast_{a}\ast_{b}\ast_{c} h^{(l_1)}(x_1;a)h^{(l_2)}(x_2;b)
h^{(l_3)}(x_3;c)\qquad\qquad\qquad
\end{eqnarray}
where the asterisks denote the three contractions over $a,b,c$. The homogeneities in $a,b,c$ are in fact the same on both sides of these. The middle term is from (5.12).

Evaluating (5.16) we obtain a linear combination of terms such as
\begin{eqnarray}
(a\cdot b)^{Q_{12}} (b\cdot c)^{Q_{23}} (c\cdot a)^{Q_{31}}\qquad\qquad\qquad\qquad \nonumber\\ \times
\{(a\cdot\partial_1)^{k_1}(b\cdot \partial_1)^{k_2} (c\cdot \partial_1)^{k_3}
\Box_1^{q}(a\cdot \partial_2)^{k'_1}(b\cdot \partial_2)^{k'_2}(c\cdot \partial_2)^{k'_3} \Box_2^{q'}
\qquad \nonumber\\ \times (a\cdot \partial_3)^{k''_1}(b\cdot \partial_3)^{k''_2}
(c\cdot \partial_3)^{k''_3}\Box_3^{q''}\}\ast_{a} \ast_{b} \ast_{c} h^{(l_1)}(x_1;a)
h^{(l_2)}(x_2;b) h^{(l_3)}(x_3;c)
\end{eqnarray}
where the asterisks as usual demand the contractions of left and right $a,b,c$ tensors, respectively. The exponents are submitted to constraints that we have to derive now. First we count the number of derivatives. From (3.13),(5.6) we obtain the number of derivatives as $2m$
\begin{equation}
2m= 2\{\sum_{i}(p_{i}+n_{i})+ 3-D\}
\end{equation}
But each factor $x_{ij}$ compensates one derivative and the number of such
factors is from (5.4)
\begin{equation}
(p_1+p_2-l_2 +2n_1)+(p_2+p_3-l_3+2n_2) +(p_3+p_1-l_1+2n_1) =
\sum_{i}[2(p_{i}+n_{i}) - l_{i}]
\end{equation}
With
\begin{equation}
K_{i} = k_{i} +k'_{i} +k''_{i}
\end{equation}
the final number of derivatives acting on the higher spin fields is
\begin{equation}
\sum_{i}K_{i} +2(q+q'+q'') = \sum_{i} l_{i} +2(3 -D)  
\end{equation}
But the orders of $a,b,c$ to the left of the asterisks must be the same as in the higher spin fields. This gives three further constraints
\begin{eqnarray}
Q_{12}+Q_{31} +K_1 = l_1 \\
Q_{23}+Q_{12} +K_2 = l_2 \\
Q_{31}+Q_{23} +K_3 = l_3  
\end{eqnarray}
which by elimination of the $K_{i}$ from (5.21) leads to
\begin{equation}
Q_{12}+Q_{23}+Q_{31} =q+q'+q'' +(D-3)
\end{equation}
Note that if the higher spin fields are on shell, the Laplacians give $O(L^{-2})$ terms and can at leading order be skipped so that we may set $q+q'+q'' =0$. Then (5.25) gives the number of direct tensorial contractions beween pairs of higher spin fields as $D-3$. Besides the numbers $Q_{ij}$ of contractions, six parameters $k$ remain free.
 
\setcounter{equation}{0}
\section{Conclusions and outlook}
The structure of the invariants of $N$ higher-spin fields with general $N>3$
can be derived analogously. The number of derivatives acting on the fields is
\begin{equation}
\sharp \{\textnormal{derivatives}\} =\sum_{i} l_{i} +D -N(D-2)
\end{equation}
and the number of contractions between different on-shell fields is
\begin{equation}
\sharp \{\textnormal{contractions}\} = \frac{1}{2}[N(D-2) -D]
\end{equation}
According to what we said in section 5 we should identify the number of 
derivatives (6.1) with $\Delta_{max}$. The complete calculation for $N=3$ including all powers of $L^{-2}$ should be possible for small $l_{i}$ with
the help of computers yielding a basis of gauge invariant forms plus finally the corresponding gauge transformations.

\end{document}